\documentclass[12pt,preprint]{aastex}

\shorttitle{Cosmic ray pevatrons}
\shortauthors{Gabici \& Aharonian}

\begin{document}


\title{Searching for galactic cosmic ray pevatrons \\ with multi--TeV gamma rays and neutrinos}


\author{Stefano Gabici\altaffilmark{1} and Felix A. Aharonian\altaffilmark{2,1}}





\altaffiltext{1}{Max--Planck--Institut f\"ur Kernphysik, Saupfercheckweg 1, 
    69117 Heidelberg, Germany}
\altaffiltext{2}{Dublin Institute for Advanced Studies, 5 Merrion Square, Dublin 2, Ireland}


\begin{abstract}
The recent HESS detections of supernova remnant shells 
in TeV $\gamma$-rays confirm the theoretical predictions that  
supernova remnants can operate as powerful 
cosmic ray accelerators. If these objects 
are responsible for the bulk of galactic cosmic rays, 
then they should accelerate protons and nuclei to $10^{15} \rm eV$ 
and beyond, i.e. act as cosmic PeVatrons. The model of diffusive
shock acceleration allows, under certain conditions, acceleration 
of particles to such high energies and their gradual injection
into the interstellar medium, mainly during the Sedov phase of the remnant evolution.  The most energetic particles are released first,   
while particles of lower energies are more effectively confined in the shell, and 
are released at later epochs.  Thus the spectrum 
of nonthermal paticles inside the shell extends to PeV energies only
during a relatively short period of the evolution of the remnant. 
For this reason one may expect spectra of secondary $\gamma$-rays 
and neutrinos extending to energies beyond 10 TeV only from 
$T \lesssim 1000$~yr old  supernova remnants. On the other hand, if by a chance  
a massive gas cloud appears  in the $ \lesssim 100$ pc vicinity of 
the  supernova remnant, ``delayed''   
multi-TeV signals of $\gamma$-rays and neutrinos arise
when  the most energetic 
partices emerged from the supernova shell reach the
cloud.  
The detection of such delayed emission of multi-TeV
$\gamma$-rays and neutrinos allows indirect identification of the  
supernova remnant as a particle PeVatron.  
\end{abstract}





\section{Introduction}

In 1934, Baade and Zwicky proposed that supernovae are
responsible for the observed flux of cosmic rays (CR), provided 
that a substantial fraction 
of the kinetic energy 
released at explosions of 
supernovae is converted into 
relativistic particles. Later, it has been recognised that  
relativistic particles can be effectively accelerated 
via Fermi mechanism at shock waves that form during the expansion of supernova remnants (SNR) in the interstellar medium (e.g. \citealt{NLS}).

The particle acceleration in SNRs is  accompanied by production of 
$\gamma$-rays and neutrinos due to interactions of accelerated CR protons and nuclei with 
ambient medium. Recently, the HESS collaboration 
reported detection of some young SNRs in TeV $\gamma$-rays \citep{hess} with 
fluxes quite close to the early theoretical predictions \citep{dav}.
Though the HESS results constitute a very important advancements in the field, they still do not provide us with a definite and direct evidence of proton acceleration in SNRs; the  competing inverse Compton 
scattering of directly accelerated electrons may significantly contribute to the 
observed $\gamma$-ray fluxes, provided that magnetic field in the acceleration 
region does not exceed $10 \, \mu$G.  
Production of $\pi^{0}$ decay  $\gamma$-rays is accompanied by radiation of neutrinos.
Their detection, though challenging even for the next generation of telescopes \citep{halzen,kappes,neutrinos},  would provide an unambiguous evidence 
for proton acceleration in SNRs. 

The observed CR spectrum extends without any distinct feature 
up to the so called {\it knee} at an energy of a few PeV.  
This suggests that the sources of galactic CRs, whichever they are, must be able to accelerate particles up to {\it at least} a few PeV. The 
standard linear treatment of diffusive shock acceleration 
does not allow  acceleration  of protons  beyond 
$\approx 10^{14} (B/3 \mu G)$ eV \citep{lc}
which falls short of the position of the knee, unless one assumes 
strong amplification of the magnetic field in the  upstream region.  
Remarkably, the background magnetic field close to a shock 
which is efficiently accelerating particles can be amplified by a 
factor up to $\sim 10^3$ due to CR--driven instabilities \cite{bell}. The magnetic field amplification hypothesis is supported also by observations of thin X--ray synchrotron 
filaments surrounding SNR \citep{bamba,vink,vbk}.

A decisive and unambiguous indication of acceleration of PeV protons 
in SNRs can be provided by observations of $\gamma$-rays at energies 
up to 100 TeV and beyond. Because of the Klein-Nishina effect 
the efficiency of 
inverse Compton scattering 
in this energy band is dramatically reduced. Therefore unlike 
other energy intervals, the interpretation of 
$\gamma$-ray observations at these energies is  
free of confusion and reduces to the only possible 
mechanism - decay of secondary $\pi^0$-meson. 
Although the potential of the current ground-based 
instruments for detection of such energetic 
$\gamma$-rays is limited, it is expected that 
the next generation arrays of imaging Cherenkov telescopes 
optimized in the multi--TeV energy range will become 
powerful tools for this kind of studies \citep{IACT}. 

It should be noted that the number of SNRs currently bright in $> 10$ TeV $\gamma$-rays 
is expected to be rather limited. Multi--PeV protons can be 
accelerated only during a relatively short period of the SNR evolution, namely, at the end of the free--expansion phase/beginning of the Sedov phase, when the shock velocity is high enough to allow sufficiently high acceleration rate. When the SNR enters the Sedov phase, the shock gradually slows down and correspondingly the maximum energy of the particles that can be confined within the SNR decreases. This determines the escape of the most energetic particles from the SNR \citep{ptuskin}. Thus, unless our theoretical understanding of particle acceleration at SNR is completely wrong, we should expect an energy  spectrum of CR inside the SNR approaching  
PeV energies only at the beginning of the Sedov phase, typically 
for a time $\lesssim 1000$ years. 
When the remnant enters the Sedov phase, the high energy cutoffs in the spectra of 
both protons and $\gamma$-rays gradually moves to lower energies, while the highest energy particles 
leave the remnant \citep{ptuskin}. 
This can naturally explain why the $\gamma$-ray spectrum of the best studied SNR 
RX J1713.7--3946 above 10 TeV becomes rather steep with photon index $\approx 3$
\citep{hess100TeV}.

In this Letter we suggest a search for multi--TeV $\gamma$-rays 
generated by the CRs that escape the SNR. 
A molecular cloud located close to the SNR can provide an effective 
target for production of $\gamma$-rays \citep{felixclouds,atoyan,clouds}. 
The highest energy particles ($\sim$ few PeV) escape the shell 
first. Moreover, generally they diffuse in the interstellar medium 
faster than low energy particles. Therefore they arrive first to the cloud,
producing there $\gamma$-rays and neutrinos with very hard energy spectra.  
Note that an association of SNRs with clouds is naturally 
expected, especially in star forming regions \citep{montmerle}. 
The duration of $\gamma$-ray emission in this case is determined by the time 
of propagation of CRs from the SNR to the cloud.
Therefore $\gamma$-ray emission of the cloud lasts much longer
than the emission of the SNR itself. This makes the detection of 
delayed $\gamma$-ray and neutrino signals more probable.
The detection of these multi--TeV $\gamma$-rays from 
nearby clouds would thus indicate that the nearby SNR 
in the past was acting as an effective PeVatron.

\section{The model}

Consider a supernova of total energy $10^{51} E_{51} \, {\rm erg}$ exploding in a medium of density $n$. 
The initial shock velocity is $10^9 u_9$ cm/s
and remains roughly constant until the mass of the swept up material equals the mass of the ejecta.
This happens at a time  $\approx 200 [E_{51}/(n \, u_9^5)]^{1/3} {\rm yr}$, when the shock radius is $\approx 2.1 [E_{51}/(n \, u_9^2)]^{1/3} {\rm pc}$.
Then the SNR enters the Sedov phase 
and the shock radius and velocity scales with time as 
$R_{sh} \propto t^{2/5}$ and $u_{sh} \propto t^{-3/5}$.

The spectrum of particles accelerated at the SNR shock is determined by the transport equation:
\begin{equation}
\label{eq:transport}
\frac{\partial f}{\partial t} - \nabla D \nabla f + {\bf u} \nabla f - \frac{\nabla {\bf u}}{3} p \frac{\partial f}{\partial p} = 0 .
\end{equation}
where $D = D(p)$ is the momentum dependent diffusion coefficient and ${\bf u}$ the flow velocity.
For a strong shock with compression factor $r_s = 4$, the test particle theory predicts an universal shape for the CR spectrum at the shock $f_0(p) \propto p^{-4}$ \citep{NLS}. If the shock is an efficient accelerator (as SNR shocks are believed to be), the CR pressure modifies the flow structure, making the shock more compressible and the spectrum of the accelerated particles harder, $f_0(p) \propto p^{-\alpha}$ with $3.5 \lesssim \alpha \leq 4$ \citep{NLS}.  
Detailed calculations compared with multiwavelength observations of SNR suggest the values $r_s \approx 7$ and $\alpha \approx 3.7$ \citep{don,heinz}, which we adopt in the following.

The maximum momentum of the accelerated particles is determined by a simple confinement condition, namely that 
the diffusion length $l_d$ of the particles cannot exceed the characteristic size of the system $R_{sh}$: 
\begin{equation}
\label{eq:pmax}
l_d = \frac{D(p_{max})}{u_{sh}} \lesssim R_{sh} .
\end{equation}
The maximum possible energies are achieved when the acceleration proceeds in the Bohm diffusion limit, $D \propto p/B_{sh}$, with $B_{sh}$ the magnetic field strength at the shock. In this case the maximum momentum decreases with time as $p_{max}(t) \propto B_{sh} t^{-1/5}$. In fact, the drop of $p_{max}$ is even faster, given that the magnetic field is also expected to decrease with time.
This implies that at any time, particles with momentum above $p_{max}(t)$ quickly escape the remnant, generating a cutoff in the spectrum. The spectrum of the runaway particles can be approximated as a $\delta$--function (see \citealt{ptuskin}):
$$
q_{esc}(p,t) = -\delta(p-p_{max})
$$
\begin{equation}
\times \int_0^{\infty} d^3 R \left( \frac{\partial p_{max}}{\partial t} 
+ \frac{\nabla {\bf u}}{3} p_{max} \right)
f(p_{max},R)
\end{equation}
where the integration has to be performed where the integrand is negative.
Thus, to calculate the flux of the runaway particles one has to know: \textit{(i)} the CR particle distribution function at $p_{max}$ at any location in the SNR,
\textit{(ii)} the flow velocity both inside the shock and outside it, where the CR precursor forms and \textit{(iii)} how the maximum momentum varies during the SNR evolution.
\citet{ptuskin} showed that it is straigthforward to derive \textit{(i)} and \textit{(ii)} using an approximate (but reasonably accurate) linear velocity profile inside the SNR 
and assuming that the CR pressure at the shock $P^{CR}_{sh}$ is a fraction $\xi_{CR}$ of the incoming ram pressure $\varrho u_{sh}^2$ and that $f_0(p_{max}) \propto P^{CR}_{sh}$.

The determination of $p_{max}$ and its evolution with time requires the knowledge of the diffusion coefficient (see Eq. \ref{eq:pmax}), which is in turn determined by the level of magnetic turbulence generated by accelerated particles themselves. This makes the problem nonlinear and very difficult to be solved. The value of $p_{max}$ depends on a few crucial but poorly known aspects of the problem, including the nature of CR--driven instability operating in the shock precursor and the level of wave damping \citep{bell,ptuskin,pasquale,vladimirov}. 
Because of these uncertainties, we adopt here a phenomenological approach, namely we parametrize the maximum momentum as $p_{max}(t) \propto t^{-\delta}$. We further assume $p_{max} \sim 5$ PeV and $\sim 10$ GeV at the early and late epochs of the Sedov phase respectively. This requires $\delta \approx 2.4$. 
Remarkably, if the maximum momentum is a power law function of time, the spectrum of the escaping particles integrated over the whole Sedov phase is also a power law of the form $\propto p^{-4}$ \citep{ptuskin}, which is close (sligthly harder) to what needed to fit the CR data below the knee \citep{CRbook}.

The spectrum of the CRs inside the SNR $f_{in}(R,p,t)$ can be obtained from Eq. \ref{eq:transport} by dropping the diffusion term, while the spectrum of the runaway CRs at a given distance $R$ from the SNR and at a given time $t$ can be obtained by solving the diffusion equation:
\begin{equation}
\frac{\partial f_{out}}{\partial t}(R,p,t) = D_{ISM}(p) \nabla^2 f_{out}(R,p,t) + q_{esc}(p,t) \delta(\mathbf{R}) .
\end{equation}
The diffusion coefficient $D_{ISM}(p)$ describes the propagation of CRs in the galactic disk. The available CR data require a power--law energy dependence, $D_{ISM}(E) \propto E^{-s}$, with $D_{ISM} \approx 10^{28} {\rm cm}^2/{\rm s}$ at $E \approx 10$ GeV and $s \approx 0.3 \div 0.7$ \citep{CRbook}.
The constraints on the diffusion coefficient are obtained from the comparison between diffusion models and CR data and have to be considered as average galactic values. However, the conditions might be rather different in regions close to CR sources, in particular due to the presence of strong gradients in the CR distribution, which may enhance the generation of plasma waves and thus suppress the diffusion coefficient \citep{wentzel}. Below we assume $D_{ISM} = 3 \times 10^{29} (E/1 PeV)^{0.5} {\rm cm}^2/{\rm s} $.  
The change of $s$ within the allowed range or the choice of a different normalization for $D_{ISM}$ does not alter qualitatively the results, the main effect being that the characteristic time scales of the problem change prortional to $1/D_{ISM}$.

The functions $f_{in}$ and $f_{out}$ can be used to evaluate the $\gamma$-ray and neutrino fluxes due to CR interactions in the ambient gas, both from the SNR itself and surrounding dense environments (e.g. from nearby molecular clouds).

\section{Results}

The top panel of Fig. \ref{fig:1} shows the energy spectrum of $\gamma$-ray emission from interactions o accelerated protons with ambient medium, calculated for typical parameters characterizing SNRs: $E_{51} = n = u_9 = 1$. The bottom panel shows the emission from a cloud of mass $M_{cl} = 10^4 M_{\odot}$ located at a distance $d_{cl} = 100$ pc away from the SNR. 
Spectra have been calculated following \citet{kelner}. 
The distance of the SNR is assumed $D = 1$ kpc and different curves refer to different times after the supernova explosion. The efficiency of CR acceleration at the SNR shock is regulated by the parameter $\xi_{CR}$ (the ratio between the CR pressure at the shock to the shock ram pressure), which is assumed to be equal to 0.3 and constant during the SNR evolution. This assumption is reasonable for strong shocks, for which the acceleration efficiency saturates to very high values \citep{ioNLS}, and it becomes less reliable in the late stages of the Sedov phase, when the SNR shock becomes progressively weaker.

Early in the Sedov phase (curve 1, 400 yr after the explosion), the $\gamma$-ray spectrum from the SNR is hard and extends up to $\gtrsim 100$ TeV, revealing the acceleration of PeV particles. The hardness of the spectrum reflects the fact that, due to nonlinear effects in particle acceleration, the underlying CR spectrum becomes harder than $p^2 f_0(p) \propto p^{-2}$. Conversely, the $\gamma$-ray flux from the cloud is extremely weak, because for the epoch of 400 yr after the explosion CRs do not have sufficient time to reach the cloud. The emission of $\gtrsim 100 {\rm TeV}$ photons from the SNR lasts a few hundreds years, and after that the cutoff in the $\gamma$-ray spectrum moves to lower energies (curves 2, 3 and 4 correspond to the epochs of $2 \, 10^3$, $8 \, 10^3$ and $3.2 \, 10^4$ yr after the explosion).
As time passes, CRs finally reach the cloud and produce there $\gamma$-rays when interacting with the dense cloud environment. This makes the cloud an effective multi--TeV $\gamma$-ray emitter, with a flux at the sensitivity level of next generation Cherenkov telescopes operating in that energy range. 
As lower and lower energy particles reach the cloud, the peak of the $\gamma$-ray emission accordingly shifts towards the lower, TeV and GeV, energies at flux levels which can be probed by ground based instruments and GLAST. 

The shape of the $\gamma$-ray spectrum is naturally explained as follows: at a time $t$, only particles with energy above $E_{*}$, given by $d_{cl} \approx \sqrt{6 D_{ISM}(E_{*}) t}$, reach the cloud. Thus the CR spectrum inside the cloud has a sharp low energy cutoff at $E_{*}$.
The corresponding $\gamma$-ray spectrum peaks at the energy $\approx 0.1 E_{*}$ and the spectral slopes above and below the peak are $\sim E^{-2-s}$ and $\sim E^{-1}$ respectively.

The multi-TeV hadronic $\gamma$-ray emission from the cloud is significantly weaker than the one from the SNR, but its detection might be easier because of its longer duration ($\lesssim 10^4$ yr versus few hundreds years). Moreover, 
the leptonic contribution to the cloud emission is likely to be negligible.
Electrons accelerated at the SNR cannot reach the cloud because they remain confined in the SNR due to severe synchrotron losses. Secondary electrons can be produced in the cloud, but they cool mainly via synchrotron emission in the cloud megnatic field $\sim 10 \div 100 \mu G$ \citep{crutcher}. This makes the production of $\gtrsim$ TeV $\gamma$-rays due to inverse Compton scattering and non--thermal Bremsstrahlung negligible.

Fig. \ref{fig:2} shows the muonic neutrino fluxes from the SNR (top panel) and the cloud (bottom panel) for the same parameters adopted in Fig. \ref{fig:1}. The flux at Earth is a factor of $\approx 2$ smaller than what showed due to neutrino oscillations. For a young SNR (curve 1), the spectrum extends up to $\sim 100$ TeV, where km$^3$--scale neutrino telescopes achieve their best performance. The flux level is $\gtrsim 10^{-11} {\rm TeV} {\rm cm}^{-2} {\rm s}^{-1}$, which makes such sources detectable in a few years. Unfortunately, the high energy cutoff in the neutrino spectrum moves fast towards low energies (curves 2 and 3), making the detection more problematic.
Note that the $\gamma$-ray and neutrino fluxes from the SNR itself and from the cloud scale as $n/D^2$ and $M_{cl}/D^2$ respectively.
Thus, the presence of a very massive molecular cloud with mass $M_{cl} \sim 10^5 M_{\odot}$ would considerably increase the chances of detection of multi-TeV $\gamma$-rays and neutrinos from such systems.

\begin{figure}
\plotone{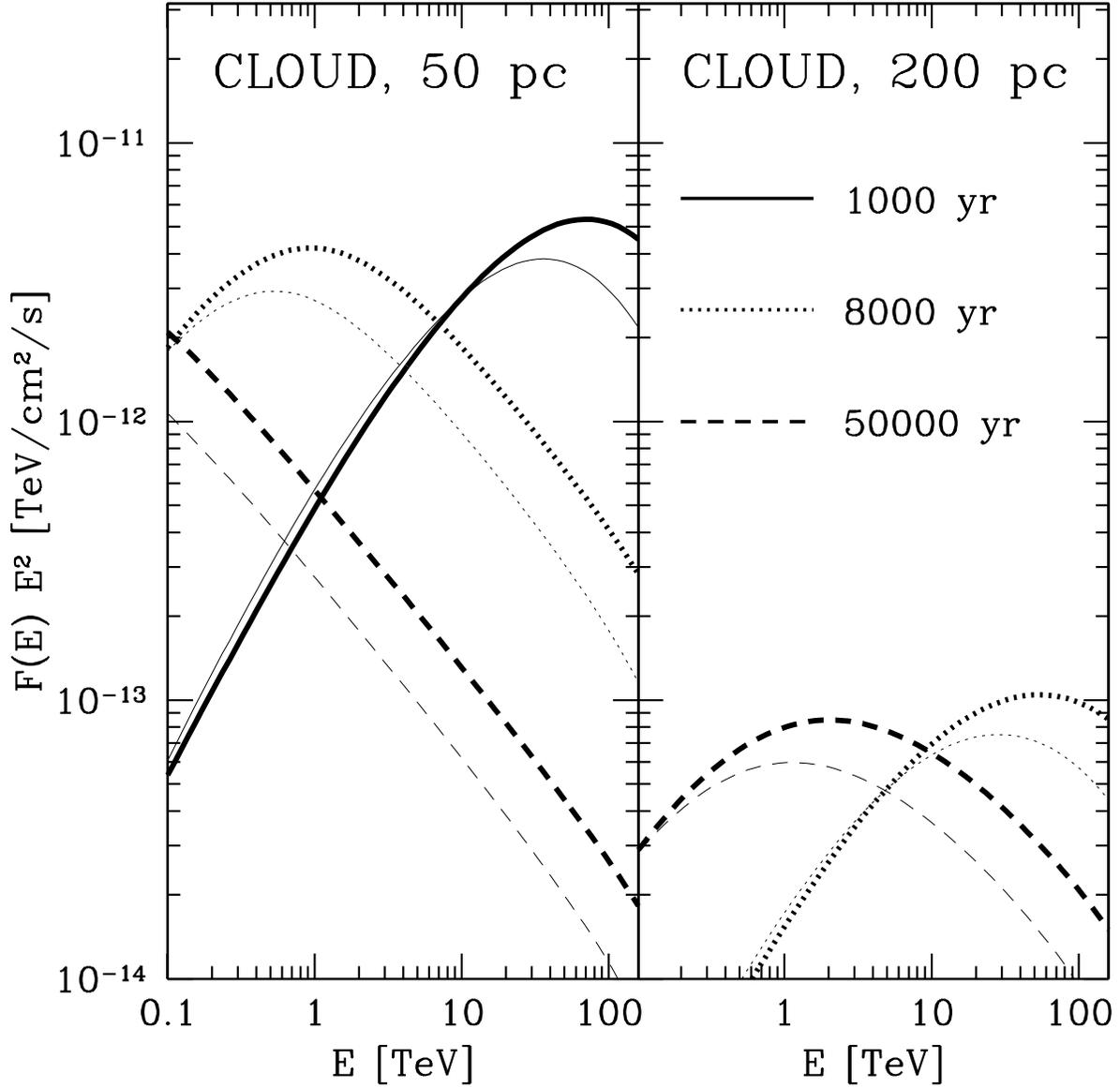}
\caption{Gamma ray (thick lines) and neutrino (thin lines) spectra from a cloud located at 50 (left) and 200 pc (right) from the SNR. Different curves refer to different times after the explosion.}
\label{fig:3}
\end{figure}

The impact of the distance between the SNR and the cloud on the flux of $\gamma$-ray and neutrino emission is illustrated in Fig. \ref{fig:3}, where $\gamma$-ray (thick lines) and neutrino (thin lines) spectra are shown for a cloud at a distance of 50 and 200 pc from the SNR. Note that the neutrino flux of the cloud may become marginally detectable only if the cloud is very close to the SNR. On the other hand, clouds can be detectable in $\gamma$-rays even if their distance from the SNR is as large as $\sim 200$ pc (right panel). Remarkably, the TeV emission from the cloud lasts $\gtrsim 10^4$ yr, significantly longer than the emission from the SNR.


\section{Conclusions}

The acceleration of CRs up to the knee in SNRs can be revealed by means of observations of multi-TeV $\gamma$-rays and neutrinos coming from the SNR and nearby molecular clouds. The emission from the clouds is weaker than the one from the SNR, but lasts longer, enhancing the probability of detection.
Both $\gamma$-rays and neutrinos are emitted with fluxes detectable by currently operating and forthcoming instruments.  
Since the $\gamma$-ray spectra from clouds are extremely  hard, $\gamma$-ray telescopes operating at very high energies ($\gtrsim 10$ TeV) would be the best instruments for this kind of study. 
A detection of such emission would both reveal the acceleration of PeV CRs and suggest the best targets for neutrino observations.

\begin{figure}
\plotone{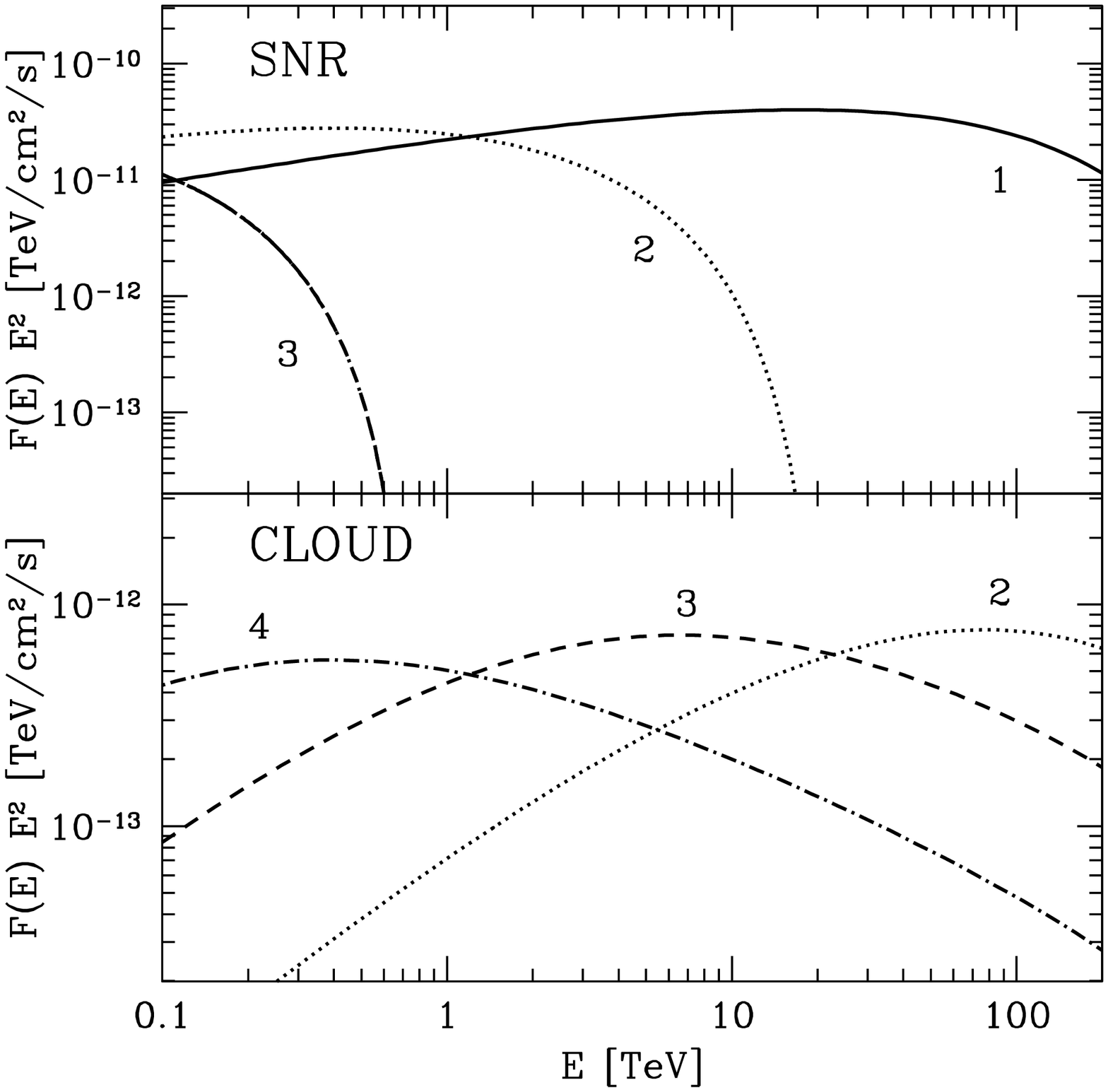}
\caption{Gamma ray spectra from the SNR (TOP) and from a cloud of $10^4 M_{\odot}$ located 100 pc away from the SNR (BOTTOM). The distance is 1 kpc. Curves refer to different times after the explosion: 400 (curve 1), 2000 (2), 8000 (3), $3.2 \, 10^4$ (4) yr.}
\label{fig:1}
\end{figure}

\begin{figure}
\plotone{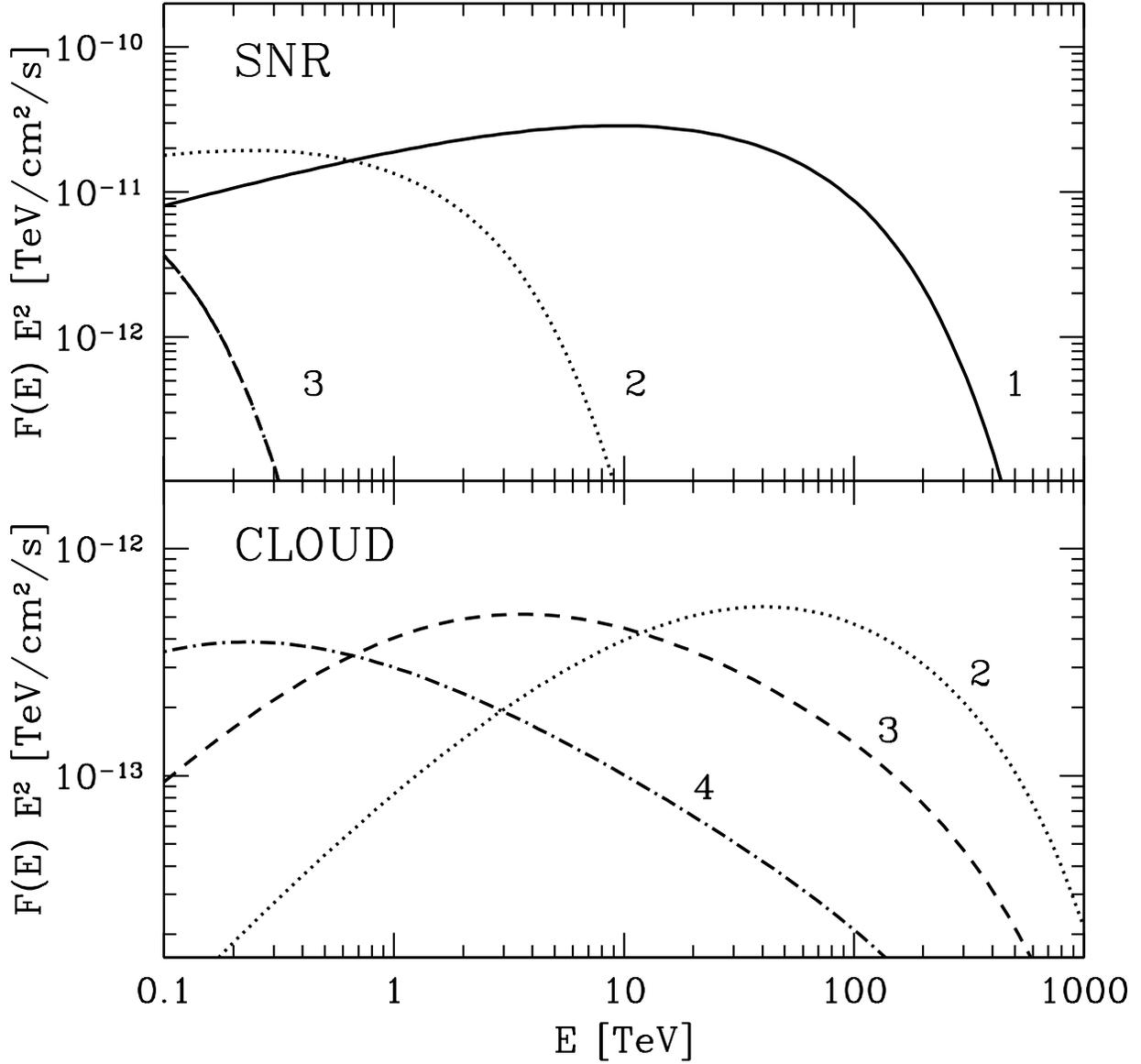}
\caption{Muonic neutrino spectra from the SNR (TOP) and from the cloud (BOTTOM) for the same set of parameters of Fig. \ref{fig:1}. Neutrino oscillations are not taken into account.}
\label{fig:2}
\end{figure}

Finally, we note that \citet{bamba2} recently proposed an association between unidentified TeV sources 
\citep{hegra,unid,MILAGRO} and old SNRs. Such idea seems to us questionable due to the fact that old SNRs cannot confine multi-TeV particles and thus emit effectively $\gamma$-rays only at sub-TeV energies (see Fig. \ref{fig:1}). In the scenario we proposed in this paper, unidentified TeV sources might still be associated with old SNRs, the gamma ray emission being produced in the interaction of escaping CRs with nearby dense clouds.



\acknowledgments

\clearpage

\clearpage

\clearpage

\end{document}